\documentclass[10pt,a4paper,twoside,final]{article}

\usepackage[a4paper,
		margin=2cm,
		top=2.5cm,
		bottom=1cm,
		textwidth=183mm,
		includefoot,
		foot=1cm,
		head=2cm]{geometry}
		
\usepackage{amsmath,amsfonts, amssymb, amsthm, wasysym}
\usepackage{xcolor, framed, graphicx}
\usepackage{url}
\usepackage{hyperref}
\usepackage{sidecap}
\usepackage[font=small,labelfont=bf]{caption}

\usepackage{siunitx}

\newcommand{\E}[1]{\bar{#1}}
\newcommand{\Var}[1]{\sigma^2_{#1}}
\newcommand{\Std}[1]{\sigma_{#1}}
\newcommand{\autocite}{\cite}
\newcommand{\textcite}{\cite}
\newcommand{\citep}{\cite}
\newcommand{\citet}{\cite}

\title{Dynamic Adaptive Computation: Tuning network states to task requirements}

\author{
Jens Wilting\,$^{1}$, Jonas Dehning\,$^{1}$, Joao Pinheiro Neto\,$^{1}$, Lucas Rudelt\,$^{1}$, Michael Wibral\,$^{2}$,\\ Johannes Zierenberg\,$^{1,3}$ and Viola Priesemann\,$^{1,3,*}$ \\
{\small $^{1}$Max-Planck-Institute for Dynamics and Self-Organization, G\"ottingen, Germany} \\
{\small$^{2}$Magnetoencephalography Unit, Brain Imaging Center, Johann-Wolfgang-Goethe University, Frankfurt, Germany } \\
{\small$^{3}$Bernstein-Center for Computational Neuroscience, G\"ottingen, Germany }
}
\date{ \normalsize \today}

\begin{document}

\maketitle

\begin{abstract}

Neural circuits are able to perform computations under very diverse conditions and requirements.
The required computations impose clear constraints on their fine-tuning: a rapid and maximally informative response to stimuli in general requires decorrelated baseline neural activity. Such network dynamics is known as asynchronous-irregular. In contrast, spatio-temporal integration of information requires maintenance and transfer of stimulus information over extended time periods.
This can be realized at criticality, a phase transition where correlations, sensitivity and integration time diverge. 
Being able to flexibly switch, or even combine the above properties in a task-dependent manner would present a clear functional advantage.
We propose that cortex operates in a “reverberating regime” because it is particularly favorable for ready adaptation of computational properties to context and task.
This reverberating regime enables cortical networks to interpolate between the asynchronous-irregular and the critical state by small changes in effective synaptic strength or excitation-inhibition ratio. These changes directly adapt computational properties, including sensitivity, amplification, integration time and correlation length within the local network.
We review recent converging evidence that cortex \textit{in vivo} operates in the reverberating regime, and that various cortical areas have adapted their integration times to processing requirements.
In addition, we propose that neuromodulation enables a fine-tuning of the network, so that local circuits can either decorrelate or integrate, and quench or maintain their input depending on task.
We argue that this task-dependent tuning, which we call ``dynamic adaptive computation'', presents a central organization principle of cortical networks and discuss first experimental evidence.

\end{abstract}

\vspace*{1cm}

\begin{figure}
\includegraphics[width=180mm]{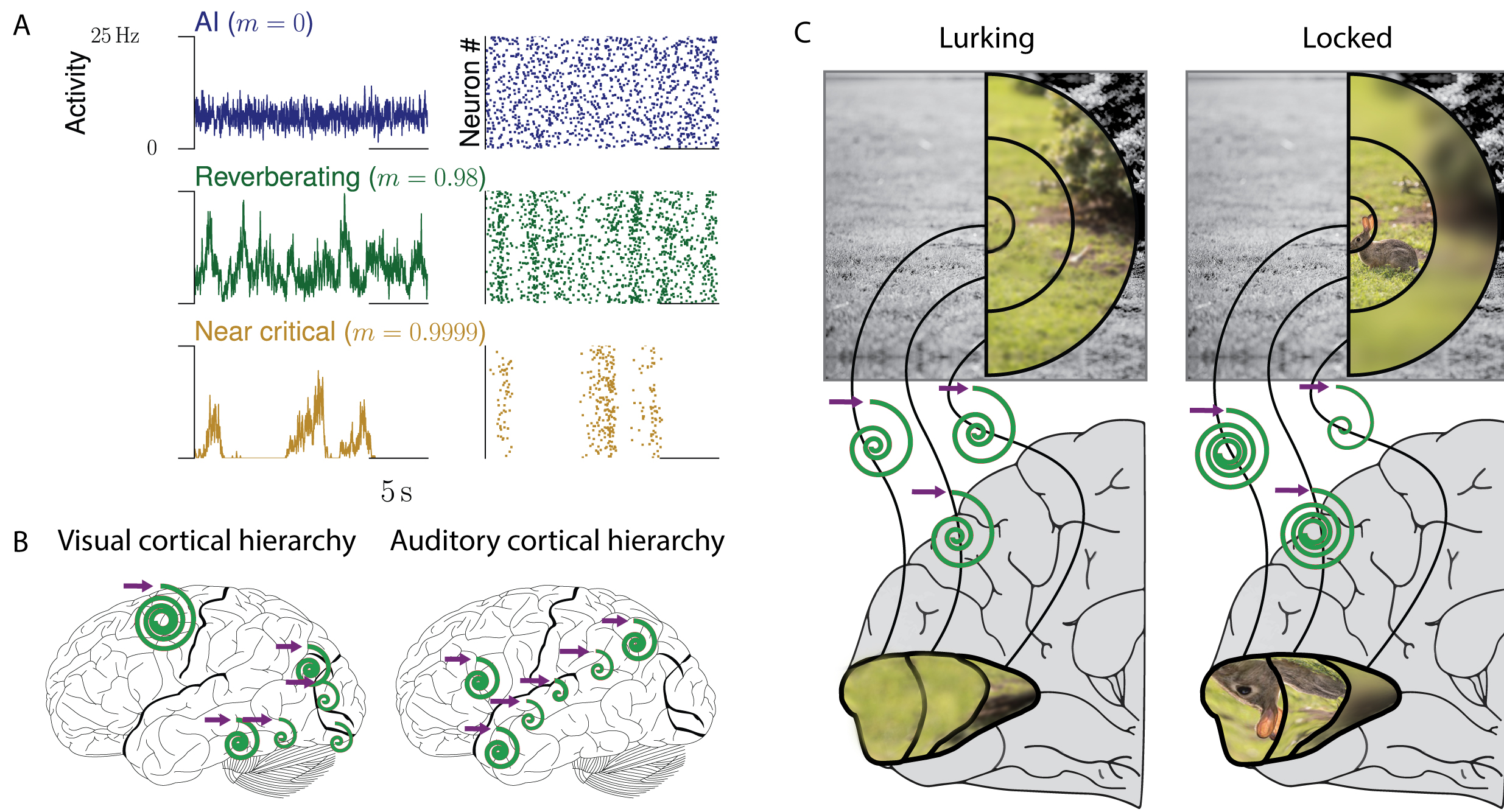}
\caption{
\textbf{Collective dynamics of cortical networks.}
\textbf{A.} Examples of collective spiking dynamics representing either irregular and uncorrelated activity (blue), reverberations (green), or dynamics close to a critical state (yellow). Population spiking activity $a(t)$ and raster plots of 50 neurons are shown. 
\textbf{B.} Hierarchical organization of collective cortical dynamics. In primary sensory areas, input is maintained and integrated only for tens of milliseconds, whereas higher areas show longer reverberations and integration. The purple arrow represents any input to the respective area, the spirals the maintenance of the input over time (inspired from \citep{Hasson2015}). Illustration adapted from \url{https://commons.wikimedia.org/wiki/File:Human-brain.SVG} (attribution: Hugh Guiney) under CC BY-SA.
\textbf{C.} Dynamic adaptation of collective dynamics in local circuits. When a predator is lurking for prey, the whole field of view needs to be presented equally in cortex. Upon locking on prey, attention focuses on the prey. This could be realized by local adaptation of the network dynamics, which amplifies the inputs from the receptive fields representing the rabbit (``tuning in''), while quenching others (``tuning out'').
Illustration adapted from \url{https://commons.wikimedia.org/wiki/File:Retinotopic_organization.png} under CC BY-SA.
}
\label{fig:1}
\end{figure}

Cortical networks are confronted with ever-changing conditions, whether these are imposed on them by a natural environment, or induced by the actions of the subjects themselves.
For example, when a predator is lurking for a prey it should detect the smallest movement in the bushes anywhere in the visual field, but as soon as the prey is in full view and the predator moves to strike, visual attention should focus on the prey (Fig. \ref{fig:1}A).
Optimal adaptation for these changing tasks requires a precise and flexible adjustment of input amplification and other properties within the local, specialized circuits of primary visual cortex: strong amplification of small input while lurking, but quenching of any irrelevant input when chasing.
These are changes from one task to another.
However, even the processing within a single task may require the joint contributions of networks with diverse computational properties.
For example, listening to spoken language involves the integration of phonemes at the timescale of milliseconds to words and whole sentences lasting for seconds.
Such temporal integration might be realized by a hierarchy of temporal receptive fields, a prime example of adaption to different processing requirements of each brain area \citep[Fig. \ref{fig:1}B]{Murray2014a, Hasson2015}.

Basic network properties like sensitivity, amplification, and integration timescale optimize different aspects of computation, and hence a generic input-output relation can be used to infer signatures of the computational properties, and changes thereof \citep{Kubo1957,Wilting2018}.
Throughout this manuscript, we refer to \emph{computation} capability in the following two, high-level senses.
First, the integration timescale determines the capability to process sequential stimuli.
If small inputs are quenched away rapidly, the network may quickly be ready to process the next input.
In contrast, networks that maintain input for long timescales may be slow at responding to novel input, but instead they can integrate information and input over extended time periods \citep{Boedecker2012,DelPapa2017,Lazar2009,Bertschinger2004}.
This is at the heart of reservoir computing in echo state networks or liquid state machines  \citep{Buonomano1995,Maass2002,Jaeger2004,Schiller2005,Jaeger2007,Boedecker2012}.
Second, the detection of small stimuli relies on a sufficient amplification \citep{Douglas1995}.
However, increased sensitivity to weak stimuli can lead to increased trial-to-trial variability \citep{Gollo2017}.

These examples show that local networks that are tuned to one task may perform worse at a different one, and there is no one-type-fits-all network for every environmental and computational demand.
How does a neural network manage to both react quickly to new inputs when needed, but also maintain memory of the recent input, e.g. when a human listens to language?
Did the brain evolve a large set of specialized circuits, or did it develop a manner to fine-tune its circuits quickly to the computational needs?
A flexible tuning of response properties would be desirable in the light of resource and space constraints.
Indeed, in experiments one of the most prominent features of cortical responses is their strong dependence on cognitive state and context.
For example, the cognitive state clearly impacts the strength, delay and duration of responses, the trial-to-trial variability, the network synchrony, and the cross-correlation between units \citep{Kisley1999,Goard2009,Curto2009,Marguet2011,Scholvinck2015, Harris2011,Kohn2009,Poulet2008,Massimini2005a,Priesemann2013}.
Transitions between different cognitive states have been described by phase transitions \cite{Steyn-Ross2010,Steyn-Ross2010a,Galka2010}.

While a phase transitions can be very useful to realize cognitive state changes, we here want to emphasize a particular property of systems close to phase transitions: without actually crossing the critical  point, already small changes in $m$ can have a large impact on the network dynamics and function.
Hence a  classical phase transition may not be necessary for adaptation.
In addition to the well-established phase transitions, adaptation could be realized as a dynamic process that regulates the proximity to a phase transition and allows cortical networks to fine-tune their sensitivity, amplification, and integration timescale within one cognitive state, depending on the specific requirements.
In order to allow efficient adaptation, cortical networks must evidently satisfy the following requirements.
(i) The network properties are easily tunable to changing requirements, e.g. the required synaptic  or neural changes should be small.
(ii) The network is fully functional in its ground state, and also in the entire vicinity, i.e. the adaptive tuning does not destabilize or dysfunctionalize it.
(iii) The network receives, modifies and transfers information according to its needs, e.g. it amplifies or quenches the input depending on task.
(iv) The network's ground state in general should enable integration of input over any specific past window, as required by a given task.



We propose that cortex operates in a particular dynamic regime, the ``reverberating regime'', because in this regime small changes in neural efficacy can tune computational properties over a wide range – a mechanism that we propose to call \emph{dynamic adaptive computation}.
In this regime a cortical circuit can interpolate between two states described below, which both have been hypothesized to govern cortical dynamics and optimize different aspects of computation \citep{Burns1976, Softky1993,Stein2005,Vreeswijk1996a,Brunel2000,Beggs2003,Beggs2012,Plenz2014,Tkacik2014,Humplik2017,Munoz2017,Wilting2018,Wilting2018b}.

In the following, we recapitulate the computational properties of these two states and then identify recent converging evidence that in fact the reverberating regime governs cortical dynamics \textit{in vivo}.
We then show how specifically the reverberating regime can combine the computational properties of the two extreme states while maintaining stability and thereby satisfies all requirements for cortical network function postulated above.
Finally, we outline future theoretical challenges and experimental predictions.

One hypothesis suggests that spiking statistics in the cortical ground state is asynchronous and irregular \citep{Burns1976, Softky1993,Stein2005}, i.e. neurons spike independently of each other and in a Poisson manner (Fig. \ref{fig:1}A). Such dynamics may be generated by a ``balanced state'', which is characterized by weak recurrent excitation compared to inhibition \citep{Vreeswijk1996a,Brunel2000}.
The typical balanced state minimizes redundancy, has maximal entropy in its spike patterns, and supports fast network responses \citep{Deneve2016,Vreeswijk1996a}.
The other hypothesis proposes that neuronal networks operate at criticality \citep{Bienenstock1998,Beggs2003,Levina2007,Beggs2012,Plenz2014,Tkacik2014,Humplik2017,Munoz2017,Kossio2018}, and thus in a particularly sensitive state at a phase transition.
This state is characterized by long-range correlations in space and time, and in models optimizes performance in tasks that profit from  extended reverberations of input in the network \citep{Bertschinger2004,Haldeman2005,Kinouchi2006,Wang2011b,Boedecker2012,Shew2013,DelPapa2017}. 
These two hypotheses, asynchronous-irregular and critical, can be interpreted as the two extreme points on a continuous spectrum of response properties to minimal perturbations, the first quenching any rate perturbation quickly within milliseconds, the other maintaining it for much longer. Hence the two hypotheses clearly differ already in the basic response properties they imply.

\begin{SCfigure}
\includegraphics[width=90mm]{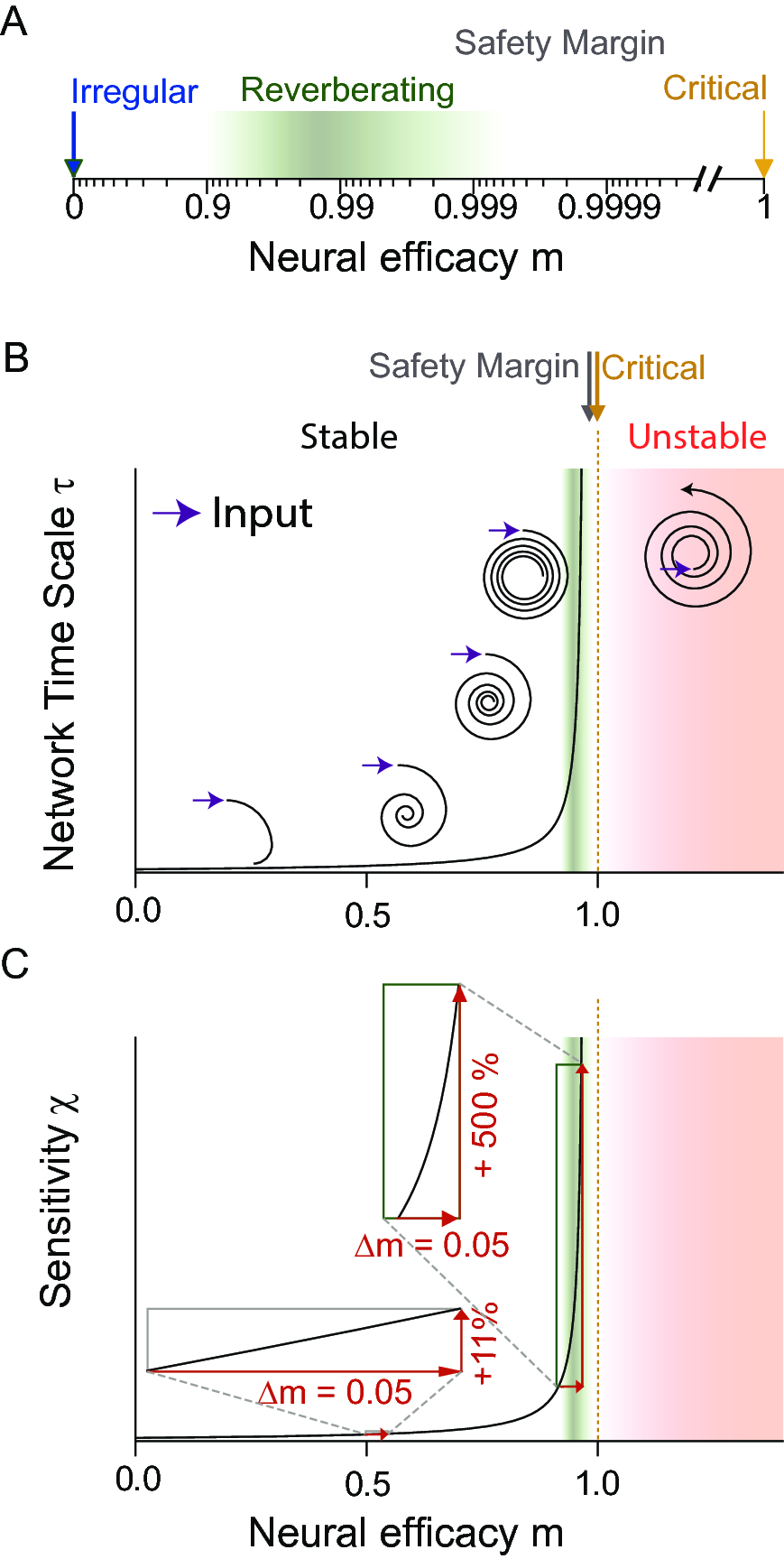}
\caption{
The neural efficacy $m$ determines the average impact any spike has on the network. Depending on $m$, network dynamics can range from irregular ($m=0$) to critical ($m=1$) and unstable ($m>1$) dynamics. 
\textbf{A.} In a logarithmic depiction of $m$, the ``reverberating regime'' (green) observed for cortex in vivo is well visible. It has clearly a larger $m$ than the irregular state (blue), but maintains a safety margin to criticality (yellow) and the instability associated with the supercritical regime (red in \textbf{B} and \textbf{C}).
\textbf{B,C.} Sketch to illustrate the divergence of dynamical and computational properties at a critical phase transition, at the example of the network timescale and the sensitivity, respectively.  \textbf{B.} The network timescale determines how long input is maintained in the network. While any rate change  is rapidly quenched close to the irregular state ($m=0$), input ``reverberates'' in the network activity for increasingly long timescales when approaching criticality ($m=1$). In the reverberating regime, the network timescale is tens to hundreds of milliseconds.
For $m>1$, input is amplified by the network, implying instability (assuming a supercritical Hopf bifurcation here for illustration). The reverberating regime keeps a sufficient safety margin from this instability.
\textbf{C.} 
The reverberating regime found \textit{in vivo} allows large tuning of the sensitivity by small changes of the neural efficacies (e.g. synaptic strength or excitation-inhibition balance), in contrast to states further away from criticality (insets).
}
\label{fig:2}
\end{SCfigure}

A general first approach to characterize the response properties of any dynamical system is based on linear response theory: When applying a minimal perturbation or stimulation, e.g. adding a single extra spike to neuron $i$, the basic response is characterized by $m_i$,  the number of \textit{additional} spikes triggered in all postsynaptic neurons \citep{London2010}, which can be interpreted as efficacy of the one neuron.
If the efficacy is sufficiently homogeneous across neurons, then the average \textit{neural efficacy} $m$ represents a control parameter, and quantifies the impact of any single extra spike in a neuron on its postsynaptic neurons, and thus the basic network response properties to small input. 
In the next step, any of these triggered spikes in turn can trigger spikes in a similar manner, and thereby the small stimulation may cascade through the network.
The network response may vary from trial to trial and from neuron to neuron, depending on  excitation-inhibition ratio, synaptic strength, and membrane potential of the postsynaptic neurons.
Thus $m$ does not describe each single response, but the expected (average) response of the network, and thereby enables an assessment of the network's stability and computational properties.
The magnitude of the neural efficacy $m$ defines two different response regimes: If one spike triggers on average less than one spike in the next time step ($m<1$), then any stimulation will die out in finite time.
For $m>1$, stimuli can be amplified infinitely, and $m=1$ marks precisely the transition between stable and unstable dynamics (Figure \ref{fig:2}B).
In addition, $m$ directly determines the amplification of the stimulus, the duration of the response, the intrinsic network timescale and the response variability, among others in the framework of autoregressive processes \citep{Wilting2018,Wilting2018b,Harris1963}. 
Although details may depend on the specific process or model, many results presented in the following are qualitatively universal across diverse models that show a (phase) transition from stable to unstable, from ordered to chaotic, or from non-oscillatory to oscillatory activity. These include for example AR(1), Kesten, branching, and Ornstein-Uhlenbeck processes, systems that show a Hopf bifurcation, or systems at the transition to chaos~\citep{Huang2017,Harris1963,Wilting2018,Wilting2018b,Camalet2000,Boedecker2012}. 
Hence the principle of dynamic computation detailed below can be implemented and exploited in very diverse types of neural networks, and thus presents a general framework.

The neural efficacy $m$ is a statistical description of the effective recurrent activation, which takes into account both excitatory and inhibitory contributions.
We here use it to focus on the mechanism of dynamic adaptive computation in an isolated setting, instead of including as many details as possible.
We take this approach for two reasons.
First, this abstraction allows to discuss possible generic principles for adaptation.
Second, our approach enables us to assess the network state $m$ and possible adaptation $m(t)$ from experiments.
The abstract adaptation principles we consider here can be implemented by numerous physiological mechanisms, including top-down attention, adaptation of synaptic strengths or neuronal excitability, dendritic processing, disinhibition, changes of the local gain, or up-and-down states \citep{Ramalingam2013,Hirsch1991,London2005,Karnani2016,Piech2013,Wilson2008}.

Inferring the neural efficacy $m$ experimentally is challenging, because only a tiny fraction of all neurons can be recorded with the required millisecond precision \citep{Priesemann2009,Ribeiro2014,Levina2017}. In fact, such spatial subsampling can lead to strong underestimation of correlations in networks, and subsequently of $m$ \citep{Wilting2018,Wilting2018b}. However, recently, a subsampling-invariant method has been developed that enables a precise quantification of $m$ even from only tens of recorded neurons \citep{Wilting2018,Wilting2018b}. 
Together with complementary approaches, either derived from the distribution of covariances or from a heuristic estimation, evidence is mounting that  $m$ is between $\approx 0.9$ and $\approx 0.995$, consistently for visual, somatosensory, motor and frontal cortices as well as hippocampus \citep{Wilting2018,Wilting2018b,Priesemann2014,Dahmen2016}. Hence, collective spiking activity is neither fully asynchronous nor critical, but  in a \textit{reverberating regime} between the two, and any input persists for tens to hundreds of milliseconds \citep{Wilting2018,Wilting2018b,Priesemann2014,Murray2014a,Dahmen2016,Hasson2015}.
In more detail, both Dahmen and colleagues as well as Wilting \& Priesemann estimated the neural efficacy to be about $m = 0.98$, ranging from about $0.9$ to $0.995$ when assessing spiking activity in various cortical areas (Figs. \ref{fig:1}A, \ref{fig:2}A). This magnitude of neural efficacy $m$ implies intrinsic timescales of tens to hundreds of milliseconds. Such intrinsic timescales were directly estimated from cortical recordings in macaque by Murray and colleagues, who identified a hierarchical organization of timescales across somatosensory, medial temporal, prefrontal, orbitofrontal, and anterior cingulate cortex - indicating that every cortical area has adapted its response properties to its role in information processing \citep{Murray2014a}.
These findings are also  in agreement with the experiments by London and colleagues, who directly probed the neural efficacy by stimulating a single neuron in barrel cortex. They found the response to last at least 50 ms in that primary sensory cortex, implying $m \gtrsim 0.92$ \citep{London2010,Wilting2018,Wilting2018b}.


We will show that this reverberating regime with median $m=0.98$ ($0.9<m<0.995$) observed \textit{in vivo} is optimal in preparing cortical networks for flexible  adaptation to a given task, and meets all the requirements for \textit{dynamic adaptive computation} we postulated above. 

The reverberating regime allows tuning of computational properties by small, physiologically plausible changes of network parameters.
This is because the closer a system is to criticality ($m=1$), the more sensitive its properties are to small changes in the neural efficacy $m$, e.g. to any change of synaptic strength or excitation-inhibition ratio (Fig.\ref{fig:2}C).
In the reverberating regime the network can draw on this sensitivity: inducing small overall synaptic changes allows to adapt network properties to task requirements over a wide range.
Assuming for example an AR(1), Ornstein-Uhlenbeck, or branching process, increasing $m$ from $0.94$ to $0.99$ leads to a six-fold increase in the sensitivity of the network  (Figure \ref{fig:2}C). 
Here, the sensitivity $\partial r / \partial h \sim (1 - m)^{-1}$ describes how the average network rate $r$ responds to changes of the the input $h$ \citep{Wilting2018}.
In contrast, the same absolute change from $m=0.5$ to $m=0.55$ only increases the sensitivity by about 11\%. Similar relations apply to the network's amplification and intrinsic timescale~\citep{Wilting2018b}.
Thereby, any mechanism that increases or decreases the overall likelihood that a spike excites a postsynaptic neuron can mediate the change in neural efficacy $m$.
Such mechanisms may act on the synaptic strength of many neurons in a given network, including neuromodulation that can change the response properties of a small population within a few hundred milliseconds, and homeostatic plasticity or long term potentiation or depression, which adapt the network over hours ~\citep{Rang2003,Turrigiano2004,Zierenberg2018}.
An alternative target could be the excitability of neurons, either rapidly by modulatory input, dendritic processing, disinhibition, changes of the local gain, or up-and-down states \citep{Larkum2013,Ramalingam2013,Hirsch1991,London2005,Karnani2016,Piech2013,Wilson2008}; or by changes of the intrinsic conductance properties over hours to days \citep{Turrigiano1994}.

While states even closer to the phase transition than $m=0.98$ would imply even stronger sensitivity to changes in $m$, being too close to the phase transition comes with the risk of crossing over to instability ($m>1$), because synapses are altered continuously by a number of processes, ranging from depression and facilitation to long-term plasticity. 
Hence, posing a system too close to a critical phase transition may lead to instabilities and potentially causes epileptic seizures \citep{Priesemann2014,Meisel2012,Wilting2018}.
In the following, we estimate that the typical synaptic variability limits the precision of network tuning, and thereby defines an optimal regime for functional tuning that is about one percent away from the phase transition.
Typically, single synapses exhibit about 50\% variability in their strengths $w$, i.e. $\Std{w} \approx 0.5 \, w$ over the course of hours and days \citep{Statman2014}.  
If these fluctuations are not strongly correlated across synapses, the variance of the single neuron efficacy $m_i \approx k \, \E{w}$ scales with the number $k$ of outgoing synapses, $\Var{m_i} \approx k \, \Var{w}$ and gives 
\begin{equation}
\sigma_{m_i} \approx 0.5 \, m_i / \sqrt{k}.
\label{eq:safetymargin}
\end{equation}
Assuming the network keeps a "safety margin" from instability ($m>1$) of three standard deviations renders the network stable 99.9\% of the time.
The remaining, transient excursions into the unstable regime may be tolerable, because even in a slightly supercritical regime ($m \gtrapprox 1$), runaway activity occurs only rarely \citep{Harris1963,Zierenberg2018}.
Thus assuming on average $k=\mathcal{O}(10,000)$ synapses per neuron \citep{DeFelipe2002} yields that a safety margin of about 1.5\% from criticality is sufficient to establish stability.
The safety margin can be even smaller if one assumes furthermore that the variability of $m_i$ among neurons in a local network is not strongly correlated, because the stability of network dynamics is determined by the average neural efficacy $m = \E{m_i}$, not by the individual efficacies. 
Furthermore, network structure might also contribute to stabilizing network activity \citep{Kaiser2010a}.
The resulting margin from criticality is compatible with the $m$ observed \textit{in vivo}.

The generic model reproduces statistical properties of networks where excitatory and inhibitory dynamics can be described by an effective excitation, e.g. because of a tight balance between excitation and inhibition \citep{Sompolinsky1988,Vreeswijk1996a,Ostojic2014,Kadmon2015,Huang2017}.
In general, models should operate in a regime with $m<1$ to maintain long-term stability and a safety-margin.
However, transient and strong stimulus-induced activation is is key for certain types of computation, such as direction selectivity, or sub- and supra-linear summation, e.g. in networks with non-saturating excitation and feedback inhibition \citep{Murphy2009,Lim2013,Hennequin2014,Rubin2015,Hennequin2017,Miller2016,Douglas1995,Suarez1995}.
These networks show transient instability (i.e. $m(t) > 1$) until inhibition stabilizes the activity.
Whether  such transient, large changes in $m(t)$ on a millisecond scale should be considered a ``state'' is an open question.
Nonetheless, experimentally a time resolved $m(t)$ can be estimated with high temporal resolution, e.g. in experiments with a trial-based design.
This measurement could then give insight into the state changes required for computation.

Besides the sensitivity, a number of other network properties also diverge or are maximized at the critical point and are hence equally tunable under dynamic adaptive computation.
They include the spatial correlation length, amplification, active information storage, trial-to-trial variability, and the intrinsic network timescale \citep{Sethna2006,Wilting2018,Barnett2013,Boedecker2012,Harris1963}.
Some of these properties, which diverge at the critical point as $(1-m)^{-\beta}$ (with a specific scaling exponent $\beta$), are advantageous for a given task; others, in contrast, may be detrimental.
For example,  at criticality the trial-to-trial variability diverges and undermines reliable responses. Moreover, in the vicinity of the critical point convergence to equilibrium slows down \citep{Scheffer2012}.
Thus, network fine-tuning most likely is not based on optimizing one single network response property alone, but represents a trade-off between desirable and detrimental aspects.
This tradeoff can be represented in the most simple case by a goal function
\begin{equation}
\Phi_\alpha = \Phi_+ - \alpha \Phi_- \propto (1 - m)^{\beta_+} - \alpha^\prime (1 - m)^{\beta_-},
\label{eq:goalfunction}
\end{equation}
which weighs the desired ($\Phi_+$) and detrimental ($\Phi_-$) aspects by a task dependent weight factor $\alpha$, and might be called free energy in the sense of \citet{Friston2010}.
Close to a phase transition, the desired and detrimental aspects diverge and depend on the critical scaling exponents $\beta_+$ and $\beta_-$.
Maximizing the goal function then yields an optimal neural efficacy $m^*$, which is here given by 
\begin{equation}
m^* = 1 - \left(\frac{\alpha^\prime \, \beta_-}{\beta_+}\right)^{-\frac{1}{\beta_+ - \beta_-}}.
\label{eq:optimum}
\end{equation}
This optimal neural efficacy (i) is in a subcritical regime unless $\alpha=0$ (i.e. detrimental aspects do not matter) and (ii) depends on the weight $\alpha$ and the exponents $\beta_+$ and $\beta_-$.
In the simplified picture of branching processes, the resulting $m^*$ determines a large set of response properties, which can thus only be varied simultaneously. 
An ``ideal'' network should combine the capability of dynamic adaptive computation with the ability to tune many response properties independently.
To which extent such a network is conceivable at all and how it would have to be designed is an open question.

One particularly important network property is the intrinsic network timescale $\tau$.
In many processes this intrinsic network timescale emerges from recurrent activation and is connected to the neural efficacy as $\tau = - \Delta t / \log(m) \approx \Delta t / (1 - m)$ \citep{Wilting2018}, where $\Delta t$ is a typical lag of spike propagation from the presynaptic to the postsynaptic neuron.
Input reverberates in the network over this timescale $\tau$, and can thereby enable short-term memory without any changes in synaptic strength (Fig. \ref{fig:2}B).
It has been proposed before that cortical computation relies on reverberating activity \citep{Herz1995a,Buonomano1995,Wang2002}.
Reverberations are also at the core of reservoir computing in echo state networks and liquid state machines \citep{Maass2002,Jaeger2007,Boedecker2012}.
Here, we extend on this concept and propose that cortical networks not only rely on reverberations, but specifically harness the reverberating regime in order to change their computational properties, in particular the specific  $\tau$, amplification, and sensitivity depending on needs.

We expect dynamic adaptive computation to fine-tune computational properties when switching from one task to the next, potentially mediated by neuromodulators, but we also expect that with development every brain area or circuit has developed computational properties that match their respective role in processing.
Experimentally, evidence for a developmental or evolutionary tuning has been provided by Murray and colleagues, who showed  that cortical areas developed a hierarchical organization as detailed above, with somatosensory areas showing fast responses ($\tau \approx 100$ms), and frontal slower ones ($\tau \approx 300$ms)  \citep{Murray2014a}.
This hierarchy indicates that the ground-state dynamics of cortical circuits is indeed precisely tuned, and it is hypothesized that the hierarchical organization provides increasingly larger windows for information integration for example across the visual hierarchy \citep{Hasson2008,Chen2015,Hasson2015,Badre2009,Chaudhuri2015}. 
In addition to that backbone of hierarchical cortical organization, dynamic adaptive computation enables the fine-tuning of a given local circuit to specific task conditions.
Indeed, experimental studies have shown that the response properties of cortical networks clearly change with task condition and cognitive state \citep{Kisley1999,Goard2009,Curto2009,Marguet2011,Scholvinck2015, Harris2011,Kohn2009,Poulet2008,Massimini2005a,Priesemann2013}.
Relating these changes to specific functional task requirements remains a theoretical and experimental challenge for the future.



\section*{Conflict of Interest Statement}
The authors declare that the research was conducted in the absence of any commercial or financial relationships that could be construed as a potential conflict of interest.

\section*{Author Contributions}

All authors were involved in the conception and revision of this study. JW, JZ, and VP wrote the manuscript. JW, JZ, and VP calculated the presented derivations. JW, JPN, and VP drafted the figures.

\section*{Funding}
All authors received support from the Max-Planck-Society.
JZ and VP received financial support from the German Ministry of Education and Research (BMBF) via the Bernstein Center for Computational Neuroscience (BCCN) Göttingen under Grant No. 01GQ1005B.
JW was financially supported by Gertrud-Reemtsma-Stiftung.
JPN received financial support from the Brazilian National Council for Scientific and Technological Development (CNPq) under grant 206891/2014-8.

\bibliographystyle{unsrt} 

\end{document}